\documentclass[pra,aps,twocolumn,secnumarabic,nobalancelastpage,amsmath,amssymb,
nofootinbib,10pt,showkeys]{revtex4-1}
\usepackage{graphics}      
\usepackage{graphicx}      
\usepackage{longtable}     
\usepackage{url}           
\usepackage{bm}            

\usepackage{lineno,hyperref}

\begin{document}


\title{Noble Gas Clusters and Nanoplasmas in High Harmonic Generation}

\author{M. Aladi, R. Bolla, P. R\'acz and I. B. F\"oldes} 

\affiliation{Wigner Research Centre for Physics of the Hungarian Academy of Sciences, Association EURATOM HAS,
H-1121 Budapest, Konkoly-Thege u. 29-33., HUNGARY}

\begin{abstract}
We report a study of high harmonic generation from noble gas clusters of xenon atoms in a gas jet. Harmonic spectra were investigated as a function of backing pressure, showing spectral shifts due to the nanoplasma electrons in the clusters. At certain value of laser intensity this process may oppose the effect of the well-known ionization-induced blueshift. In addition, these cluster-induced harmonic redshifts may give the possibility to estimate cluster density and cluster size in the laser-gas jet interaction range.
\end{abstract}


\maketitle


\section{Introduction} \label{Int}
High harmonic generation (HHG) is a promising source for the production of extreme ultraviolet, coherent radiaton in the subfemtosecond pulse duration range for decades \cite{Corkum93,Krausz09}. The most prevalent method is the use of atomic noble gases as targets. Besides atomic gases other types of media were used as well in order to increase the efficiency, i.e. molecular gases, different types of clusters and nanostructures. Noble gas clusters are possible candidates. They consist of Van der Waals bonded clumps of atoms, most often argon, krypton or xenon and they are feasible effective HHG sources \cite{Donelly96,Tisch97}. 

However the applicable highest laser intensity has an upper limit even for clusters in HHG process due to the ionization. In addition, the enhancement of the harmonic yield is observed only in a limited range of the cluster size \cite{Park14}.
On the other hand HHG spectra may give a possibility for the investigation of the nanoplasma dynamics and it can be a diagnostic tool for the study of internal cluster structure.

HHG from neutral atoms is described by the 3-step model \cite{Corkum93} in which first the electrons are tunneling out of the atom, then they are gaining energy from the laser field. When the electric field changes sign, they return to the atoms and radiate the gained energy. In case of clusters the phenomenon is more complicated as the electrons may also return to a neighbor atom or simply to the cluster. Ruf et al. \cite{Ruf13}
suggests a wave function partially delocalized over the whole cluster, to which electrons recombine coherently after tunnel ionization.

Our experiments were carried out using different noble gases for harmonics generation. In helium there is practically no cluster, xenon is the best gas for clustering, while argon is an intermediate case depending on the pressure. In the case of cluster HHG the intensity of harmonics shows a very nonlinear, steep scaling with increasing laser intensity \cite{Donelly96,Tisch97} and increasing backing pressure \cite{Tisch97} as compared with HHG intensity from atomic medium. We observed \cite{Aladi14} the steep increase of harmonics intensity with increasing pressure as a signature of clusters. This is in agreement with preliminary Rayleigh-scattering measurements.

Spectral shifts were detected in the HHG spectra depending on the laser intensity and the backing pressure \cite{Wahlstrom93}. With increasing laser intensity a frequency blueshift can be observed in the harmonics spectra both for atomic \cite{Shin01} and clustering gas \cite{Aladi14}, caused by the self-phase modulation of the propagating laser pulse. It offers the possibility to generate a continuously tuneable coherent XUV source.

At modest laser intensity however we identified a frequency shift with an opposite sign, i.e. an increasing redshift appeared with increasing backing pressure \cite{Aladi14}. In this case there are not so many free electrons in the interaction range but nanoplasmas of increasing size with increasing pressure may produce a redshift. The object of the present study is the investigation of noble gas cluster effects to the HHG process, namely the spectral redshifts of harmonics and their possible application for cluster size and density estimations.

\section{Experimental apparatus}
We used a Ti:sapphire laser system which delivers 4 mJ pulse energy and 40 fs (FWHM) pulse duration centered at a wavelength of 805 nm at 1 kHz repetition rate with an initial beam diameter of 9 mm $(1/e^2 )$. The linearly polarized laser beam was focused into the vacuum chamber using a plano-convex spherical lens of 30 cm focal length.
	
Gas jet targets were generated \cite{Aladi14,Aladi14.2} using a commercial valve (Parker series 9) with an additional nozzle which was characterized earlier using X-ray radiography to determine particle density for different pressures \cite{Rakowski05}. In our case the backing pressure $P_0$ could be varied between 1 and 12 bar. The valve had an orifice of 0.99 mm diameter and in the experiments it was opened for 1 ms duration. The nozzle had a diameter of $650\, \mu \mathrm{m}$. The experiments were carried out at room temperature in xenon gas in which it is easy to generate relatively large clusters. Note that argon results in about an order of magnitude smaller clusters compared with xenon.   
	
The frequency components of the high harmonic beam were spatially separated by a toroidal holographic grating (Jobin Yvon, 550 lines $\mathrm{mm}^{-1}$). The detector was a microchannel plate (MCP) with a phosphor screen which transforms the vacuum-ultraviolet (VUV) radiation into visible light and the screen was imaged onto a CCD detector (PCO Pixelfly). 
	
The center of the incident beam was blocked by a beam stop before the focusing optics, and in front of the grating there was an aperture to suppress the remaining of the fundamental beam \cite{Peatross94}. We obtained single shot spectra with a resolution of 0.4 nm. For comparison we chose the 25-60 nm wavelength range.
	
The distance between nozzle exit and laser beam axis was about 1 mm and the focal plane was $\sim 10$ mm in front of the nozzle exit (which means $\approx7\cdot10^{13}\,\,\mathrm{W/cm^2}$ laser intensity in the gas), i.e. the target was in the diverging beam, which preferred the short trajectory at the expense of long ones in the harmonic emission \cite{Salieres01}. In earlier experiments the intensity was varied by moving the focal plane relative to the nozzle \cite{Aladi14}. In the present series it was chosen to demonstrate the nanoplasma-caused redshift of the harmonic lines. In our arrangement the spatial filtering by the aperture preferred the short trajectories, too.

\section{Results}
\begin{figure}[ht]\centering
\includegraphics[width=\linewidth]{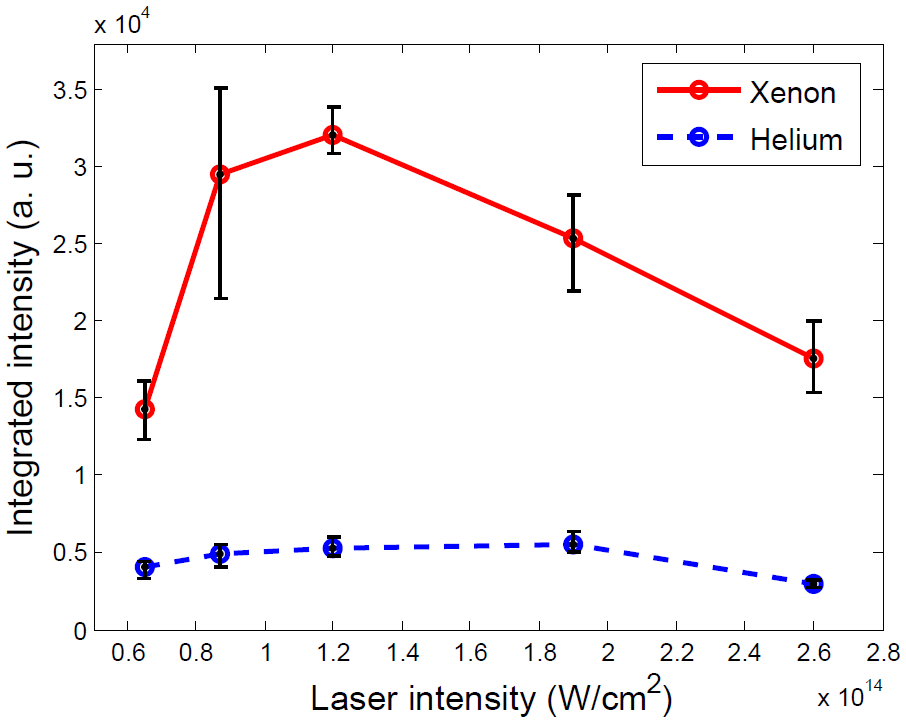}
\caption{Measured harmonic yield of the $19\,\omega$ depending on the peak laser intensity for $P_0=12$ bar in xenon and in helium.}
\label{fig1}
\end{figure}

Intensity dependence of high harmonics intensity is illustrated in Fig. \ref{fig1} for the example of $19\,\omega$ in helium and in xenon. In the case of helium a week increase with intensity is seen which saturates above $2\times10^{14}\,\mathrm{W/cm^2}$ intensity, i.e. when ionization of helium deteriorates the phase matching. Clearly, helium stays in atomic state even for the 12 bar pressure in Fig. \ref{fig1}. In contrast, xenon builds clusters easier \cite{Hagena92} and the steep increase of harmonic intensity is a typical feature of cluster building \cite{Donelly96}. On the other hand due to the lower ionization potential of xenon this sharp increase starts to saturate earlier, at $\approx1.5\times10^{14}\,\mathrm{W/cm^2}$ intensity, above which even a decrease of efficiency can be seen.
In the case of the saturation regime we observed earlier a blueshift of the harmonic lines \cite{Aladi14} as a consequence of the phase shift caused by the free electrons in the range of interaction. The appropriate intensity range for studying the nanoplasmas in the clusters is the low-intensity range in xenon, in which case the pressure dependent redshift has been observed as a signature of nanoplasmas. Therefore we carried out experiments at relatively low, $7\times10^{13}\,\,\mathrm{W/cm^2}$ laser intensity, in which case the number of free electrons is low in the medium but they are possibly present in the nanoplasmas. The spectral shift was determined for each harmonic order in dependence on the pressure. This shift can reach $\Delta\lambda_q/\lambda_q>10^{-2}$ at 12 bar. At 2 bar backing pressure clusters are already present in xenon, therefore we extrapolated the zero redshift for zero pressure.  

Using the measured redshift we can carry out cluster density and size estimations in xenon based on this nanoplasma effect.

\begin{figure}[ht]\centering
\includegraphics[width=\linewidth]{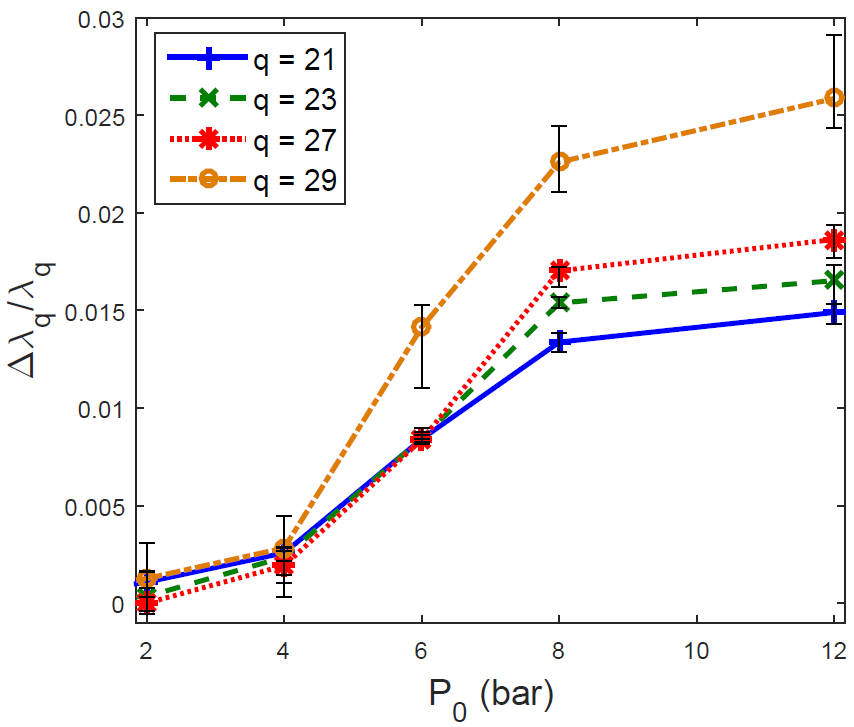}
\caption{Measured fractional spectral shifts $\Delta\lambda_q/\lambda_q $ depending on the backing pressure $P_0$ for different $q$ harmonic orders at $7\cdot10^{13}\,\,\mathrm{W/cm^2}$ laser intensity in xenon.}
\label{fig2}
\end{figure}

\begin{figure}[ht]\centering
\includegraphics[width=\linewidth]{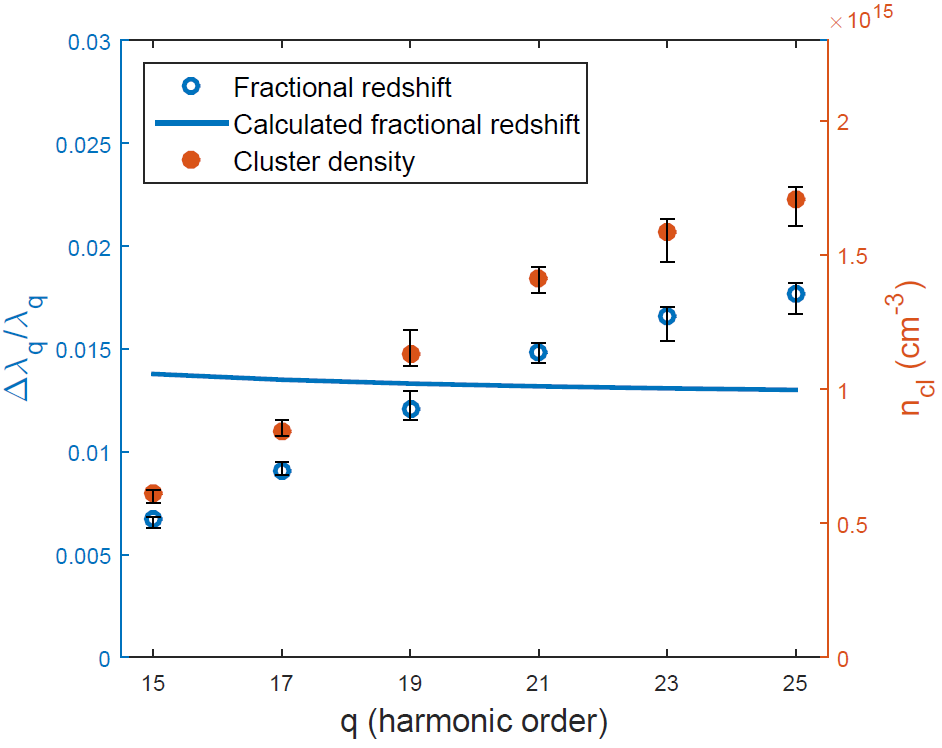}
\caption{Measured fractional spectral shift $\Delta\lambda_q/\lambda_q$ depending on the q harmonic order for $P_0=12$ bar in xenon (open circles). Calculated $\Delta\lambda_q/\lambda_q$ according to Eq. \ref{2egy} with $r_{\mathrm{cl}}=12.4$ nm and $n_{\mathrm{cl}}=10^{15}\,\mathrm{cm}^{-3}$ (solid line). Calculated cluster density from the measured $\Delta\lambda_q/\lambda_q$ values according to Eq. \ref{2egy} (solid circles).}
\label{fig3}
\end{figure}

In Fig. \ref{fig2} we can see the fractional redshifts of different harmonics $(21\,\omega-29\,\omega)$. It can be seen that the increase of the redshift with pressure starts to saturate above 8 bar. This may partly be caused by the free electron density which also grows with increasing pressure. 

The redshift dependence on the harmonic orders was investigated as well (shown in Fig. \ref{fig3}), and it was found that $\Delta\lambda/\lambda$ increases with increasing harmonic order in case of constant pressure and intensity. It is illustrated by the open circles and the left scale in Fig. \ref{fig3}. In the following section we are going to compare this $\Delta\lambda/\lambda$ with the one obtained from the nanoplasma model.

\section{Discussion}
The general phase matching condition for the qth-order harmonic is $\Delta k_{q}=k_{q}-qk_{1}=0$,
where $k_1$ is the laser vacuum wavenumber, $k_q$ is the harmonic wavenumber. Nevertheless, $\Delta k$ has several contributions in a medium originated from the Gouy phase term, the term of intensity dependent atomic dipole phase, and the dispersion term caused by the neutral gas and plasma dispersion. These contributions contribute with different signs and weights to the phase matching - both for atomic and cluster media - which should be minimal for effective harmonic generation. 
 
In cluster medium the plasma dispersion can be estimated in the frame of nanoplasma model \cite{Tisch00}. 
Whereas the blueshift of harmonics was explained \cite{Wahlstrom93} by the time-dependent phase modulation of the fundamental beam, this model is based on the dispersion relation for the laser-cluster interaction, assuming static cluster parameters, i.e. based on phase mismatch during propagation.
This model neglects the other contributions. Assuming that clusters are dielectric spheres of $r_{\mathrm{cl}}$ radius with $p$ dipole moment, the linear susceptibility of the ionized cluster medium can be calculated with the dielectric function of the nanoplasma based on Drude model ignoring electron-ion collisions. Thus the refractive index of nanoplasma medium for the $\omega$ light frequency can be written in the following form:
\begin{eqnarray}
\eta_{\mathrm{cl}}(\omega) & = & (1+4\pi\chi_{\mathrm{cl}})^{1/2}\approx 1+2\pi\chi_{\mathrm{cl}} \nonumber \\
												   & = & 1-\frac{2\pi n_\mathrm{e}r_{\mathrm{cl}}^3n_\mathrm{cl}}{3n_{\mathrm{crit}}-n_\mathrm{e}},
\label{1egy} 
\end{eqnarray}
where $n_\mathrm{e}$ is the electron density in the cluster nanoplasma, $n_{\mathrm{cl}}$ is the cluster density in the laser field, and $n_{\mathrm{crit}}$ is the critical electron density. Substituting Eq. (\ref{1egy}) into the phase mismatch expression gives the phase mismatch contribution of nanoplasma:
\begin{eqnarray}
\left(\frac{\Delta \lambda_{q}}{\lambda_q}\right)_{\mathrm{nanoplasma}}\approx -2\pi r_{\mathrm{cl}}^3 n_{\mathrm{cl}} \nonumber \\
\times\left[\frac{1}{3n_{\mathrm{crit}}/n_\mathrm{e}-1}-\frac{1}{q^{2}\,3n_{\mathrm{crit}}/n_\mathrm{e}-1}\right].
\label{2egy} 
\end{eqnarray}

\begin{figure}[ht]\centering
\includegraphics[width=\linewidth]{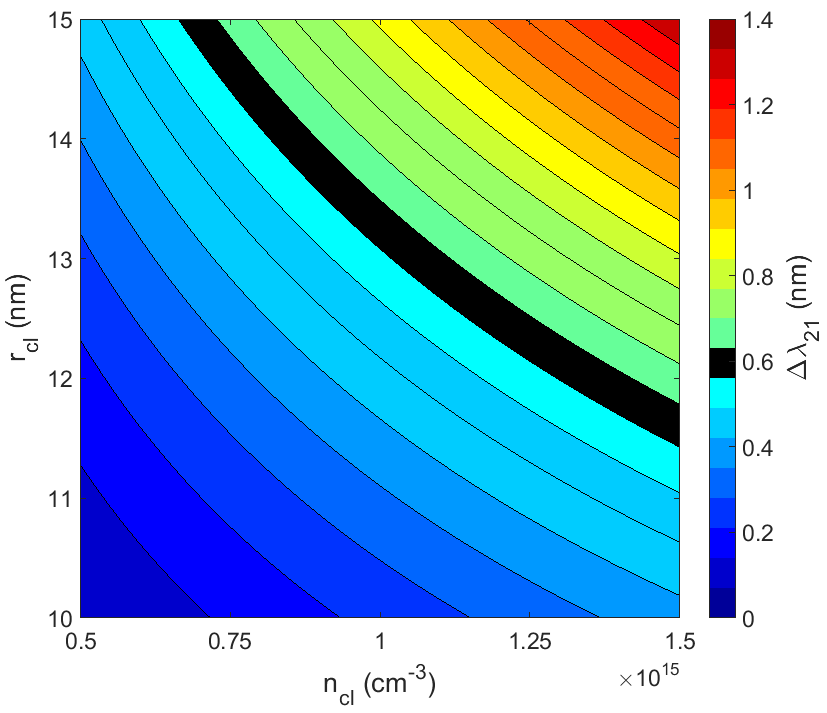}
\caption{The spectral shift $\Delta\lambda_q$ dependence on the cluster radius $r_{\mathrm{cl}}$  and the cluster density $n_{\mathrm{cl}}$ at $q=21$st order from the nanoplasma contribution according to Eq. (\ref{2egy}) in case of $n_\mathrm{e}=10^{\mathrm{23}}\, \mathrm{cm}^{-3}$ \cite{Tisch00}. The black color assigns the range of parameters for 0.6 nm wavelength shift.}
\label{fig4}
\end{figure}

These results can be compared with the measured  spectral shift of harmonics as illustrated in
Fig. \ref{fig3}. It can be seen that whereas Eq. (\ref{2egy}) gives a near constant relative wavelength shift as a function of harmonic orders, the experiments show an increase of it.   
In order to make an estimation of the cluster parameters we chose first the 21st order to compare. 
The dependence from electron density is weak, therefore we chose to compare the experimentally measured data with the function of $\Delta\lambda_q(n_{\mathrm{cl}},r_{\mathrm{cl}})$ 
with an assumed $n_e=10^{\mathrm{23}}\, \mathrm{cm}^{-3}$. This is illustrated  in Fig. \ref{fig4} for 21$\omega$. 
Thus we can compare the experimentally observed wavelength shift of a given harmonic order with the calculations of possible values of cluster density and the total cluster diameter.

The number of atoms in a cluster, $\left\langle N_{\mathrm{cl}}\right\rangle=(r_{\mathrm{cl}}/r_{\mathrm{v}})^3$. The  Van der Waals radius $r_\mathrm{v}$ is the bond distance between atoms in a cluster. Different values can be found in the literature for xenon, e.g. according to A. Bondi \cite{Bondi64} $r_\mathrm{v}=2.16\,$\,\,\AA\,, Boese et al. \cite{Boese99} $r_\mathrm{v}=2.23\,$\,\,\AA\,, but during the laser-cluster interaction clusters may expand, and larger values were used in the theoretical paper of Petrov et al. \cite{Petrov05}, $r_\mathrm{v}=2.73\,$\,\,\AA\,.

At 12 bar a $\Delta\lambda_{21}\approx0.6$ nm spectral redshift could be observed in the case of 21st order. Note that the actually measured redshift of 0.5 nm was the difference of the harmonic line between 2 and 12 bar, whereas a numerical extrapolation added an additional 0.1 nm between 2 and 0 bar backing pressure. The multiplicative factor in Eq. (\ref{2egy}) $r_{cl}^3 n_{cl}\approx n_{atom}r_{v}^3$. Thus $\left(\Delta \lambda_{q}/\lambda_q\right)\sim n_{\mathrm{atom}}r_\mathrm{v}^3$, from which the total atom density, $n_{\mathrm{atom}}$ can be expressed. We obtained for the measured $0.6$ nm redshift an atomic density of $n_{\mathrm{atom}}\approx10^{20} \,\mathrm{cm}^{-3}$ density. We can compare this value with the total gas density which was measured by x-ray shadowgraphy earlier \cite{Rakowski05} and which gave for the atomic density at 12 bar, $n_{\mathrm{atom}}\approx3\times10^{19}\,\mathrm{cm}^{-3}$. Our present result is three times higher than therein, and it is probably caused by the different applied valve which was conical therein and flat in the present experiment.

The size of the clusters in a gas jet can be estimated from the semi-empirical Hagena parameter as a guideline for the cluster size \cite{Hagena92}, too. The average number of atoms in a cluster increases with increasing backing pressure according to $\left\langle N_{\mathrm{cl}}\right\rangle \sim P_{0}^{2.35}$, however Dorchies et al. \cite{Dorchies03} gave a different, $\sim P_0^{1.8}$ scaling law. We employ the $\sim P_0^{1.8}$ scaling which is in good agreement with the scaling found in our Rayleigh-scattering measurement. Then for 12 bar backing pressure we get $\left\langle N_{\mathrm{cl}}\right\rangle\approx10^5$ and the above-mentioned range of $r_\mathrm{v}$ results in $r_{\mathrm{cl}}\approx10-13\, \mathrm{nm}$ for the cluster radii. Now we can compare it with the calculations shown in Fig. \ref{fig4} and it can be seen that the corresponding cluster density is $\sim10^{15}\, \mathrm{cm}^{-3}$ as derived from the  measured redshift. Note that larger cluster radius corresponds to smaller cluster density according to Fig. \ref{fig4}.
In Fig. \ref{fig3}  we also illustrate the estimated cluster density values for different harmonic orders with the solid circles and the right hand scale therein. The cluster radius was assumed to be constant throughout these calculations. It can be seen that although the model does not explain the obtained dependence of the relative redshifts on the order of harmonics, this causes a less than factor 2 error in the estimation of the cluster density, i.e. it can be estimated as $(1.15\pm 0.55)\times10^{15}\, \mathrm{cm}^{-3}$.

It is encouraging that the experimental data can be interpreted using the nanoplasma model, and thus estimations for the cluster size can be obtained. It must be mentioned however that according to the model the wavelength shift scales similarly to $\left\langle N_{\mathrm{cl}}\right\rangle$ with $(\Delta\lambda_q/\lambda_q)\sim P_0^{1.8}$. On the other hand however the measured pressure dependence in Fig. \ref{fig2} is different, showing a saturation at high pressures, which is probably caused by the increasing number of free electrons in the interaction range at higher pressures.
Additionally the measured $(\Delta\lambda_q/\lambda_q)$ increases with increasing harmonic order which is not expected from the calculations. 
Our result - although far from being a full diagnostics - shows that spectral redshift of the high-harmonics can be used for measuring the characteristics of nanoplasmas in clusters. For a full understanding of these processes time-dependent model for the nanoplasmas, together with the free electrons will be needed.

\section{Conclusions}
The steep pressure-dependence and the intensity-dependence of high harmonics gave already evidence of the presence of clusters in the case of xenon with high backing pressure. Additionally the observed redshift of the harmonic lines at relatively low laser intensities gives a nice demonstration of the existence of nanoplasmas in the clusters. This effect is small and the explicit detection was possible for us only at certain values of laser intensity and focus position, when the free electron density is low in the interaction domain. We showed that measuring the spectral redshift of high harmonics generated in cluster target is an applicable diagnostic method for measuring the properties of clusters. 

We analyzed the obtained results by simple model calculations which did not take into account any other factors than the nanoplasmas, but even in this case we got a relative good agreement between experiment and model calculations, and it is shown that harmonic spectral shift can be used for an order of magnitude estimation of cluster size, too. More detailed HHG experiments, i.e. measurements at even lower intensities, and improving the quality of Rayleigh-scattering diagnostics is in progress. Time-dependent simulations will be needed to understand the observed detailed pressure- and harmonic order dependence. Additionally using other clustering gases would offer an alternative possibility to understand these phenomena.   

\begin{acknowledgments}
This work has been carried out within the framework of the EUROfusion Consortium and has received funding from the Euratom research and training programme 2014-2018 under grant agreement No 633053. The views and opinions expressed herein do not necessarily reflect those of the European Commission. It was also supported by the COST MP1208 and MP1203 activities. The support and discussions with A. Czitrovszky, P. Dombi, P. Mezei and I. M\'arton is acknowledged this way.
\end{acknowledgments}


\bibliography{Ref_articles}

\end{document}